\documentclass[aps,prl,twocolumn,showpacs,superscriptaddress]{revtex4-1}

\usepackage{amsmath,amsfonts,amssymb,bm}
\usepackage{graphicx}
\usepackage{subfigure}
\usepackage{color}
\usepackage{appendix}
\usepackage[colorlinks=true, pdfstartview=FitV, linkcolor=purple, citecolor= purple, urlcolor=blue]{hyperref}

\definecolor{purple}{rgb}{0.5,0,0.5}
\definecolor{blue}{rgb}{0.0,0,0.9}

\begin{document}

\title{Revealing the Signal of QCD Phase Transition in Heavy-Ion Collisions}

\author{Yi Lu}
\email{qwertylou@pku.edu.cn}
\affiliation{Department of Physics and State Key Laboratory of Nuclear Physics and Technology, Peking University, Beijing 100871, China}

\author{Fei Gao}
\email{hiei@pku.edu.cn}
\affiliation{Center for High Energy Physics, Peking University,
	100871 Beijing, China }

\author{Xiaofeng Luo}
\email{xfluo@ccnu.edu.cn}
\affiliation{Key Laboratory of Quark \& Lepton Physics (MOE) and Institute of Particle Physics, Central China Normal University, Wuhan 430079, China}

\author{Lei Chang}
\email{{leichang@nankai.edu.cn}}
\affiliation{School of Physics, Nankai University, Tianjin 300071, China}

\author{Yu-xin Liu}
\email{yxliu@pku.edu.cn}

\affiliation{Department of Physics and State Key Laboratory of Nuclear Physics and Technology, Peking University, Beijing 100871, China}
\affiliation{Center for High Energy Physics, Peking University,
	100871 Beijing, China }
\affiliation{Collaborative Innovation Center of Quantum Matter, Beijing 100871, China}

\begin{abstract}
We propose a novel method to construct the Landau thermodynamic potential directly from the fluctuations measured in heavy-ion collisions. The potential is capable of revealing the signal of the critical end-point (CEP) and the first order phase transition (FOPT) of QCD in the system even away from the phase transition region.
With the available experimental data, we show  that the criterion  of the FOPT is   negative for most of the collision energies which indicates no signal of FOPT.  The data at $\sqrt{s_\mathrm{NN}}=7.7$ GeV with 0-5\% centrality shows a  different behavior  and the mean value of the data satisfies the criterion. However, the uncertainty is still too large to make a certain conclusion. The higher order fluctuations are also required for   confirming  the  signal.
We emphasize therefore that
new measurements with higher precision for the $C_{1,...,6}$ within 0-5\% centrality in the vicinity of $\sqrt{s_\mathrm{NN}}=7.7$ GeV  are in demand which may finally reveal the signal of  QCD phase transition.
\end{abstract}


\maketitle

{\it Introduction---}
It is of great significance to investigate the QCD phase structure for understanding the visible matter formation and  early Universe evolution~\cite{NAP13438}.
Many theoretical studies including Lattice QCD simulations have delivered solid computations at zero chemical potential and confirmed a smooth crossover with the physical quark mass~\cite{Aoki:2006,Aoki:2009,Borsanyi:2010,Bazavov:2012,Bonati:2018Tc,Bazavov:2019Tc,Borsanyi:2020Tc}.
At large chemical potential, there is still a hot debate on the existence of first order phase transition (FOPT)~\cite{Asakawa:1989,Klevansky:1992,Barducci:1994,Stephanov:1996,Alford:1998,Rapp:1998,Berges:1999}, and if exists, the location of the critical end-point~(CEP)~\cite{Fischer:2014,Gao:2016susc,Gunkel:2021,Lu:2022fz,Fu:2020,Gao:2020prd,Gao:2021plb}.
Searching for the signal of the CEP and FOPT of QCD  has then become the primary aim of relativistic heavy-ion collision (RHIC) experiments~\cite{NAP13438,STAR:2010,Mohanty:2009,Gupta:2011,Luo:2017}.

It has been proposed that the baryon number fluctuation is a possible probe for the signals~\cite{Stephanov:1999,Asakawa:2000,Hatta:2003,Stephanov:2009,Stephanov:2011,Luo:2015,Asakawa:2016}, and the corresponding technique is the beam energy scan~\cite{Luo:2015,Luo:2017}, which measures the net-proton multiplicity distribution at the chemical freeze-out line that yield the cumulants ratios, i.e. ratios between baryon number susceptibilities.
However, since the system at the freeze-out line is away from the phase transition line,
it will be difficult to verify whether the fluctuations come directly from the states at the phase boundary.
The suppression of the critical behavior due to the finite size of the fire ball further enhances the difficulty.
The difficulty is essentially because one can only measure the fluctuations at the freeze-out point,
and the signal of the FOPT could be wiped out if the system  has evolved far away from the phase transition point.
Therefore, one requires observables that are more sensitive to the order of phase transition and are capable of revealing the signal of the FOPT directly from the experimetal observations.
%

\begin{figure}[htb]
\centering
\includegraphics[width=0.42\textwidth]{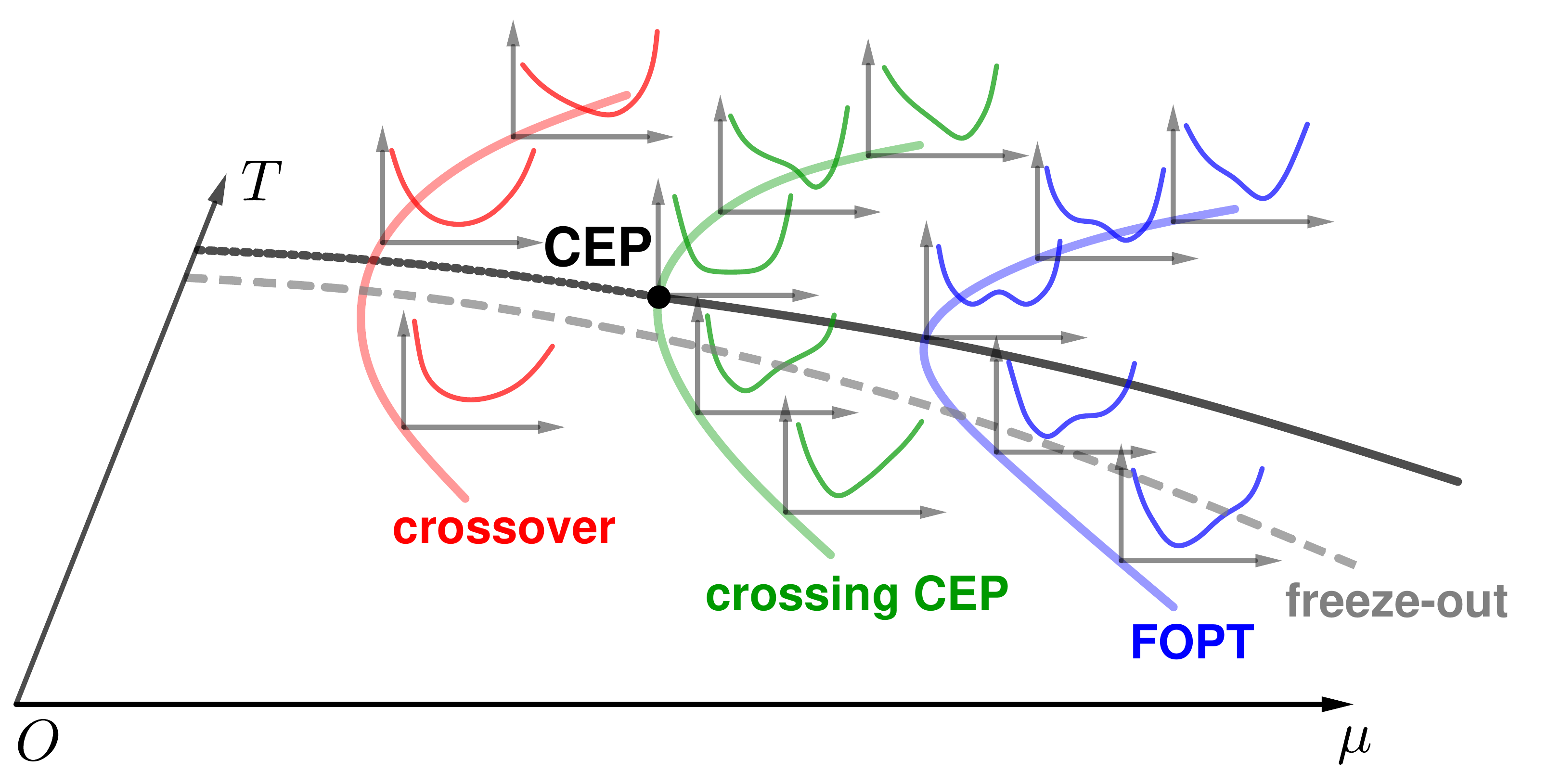}
\vspace*{-3mm}
\caption{Schematic evolution of the Landau thermodynamical potential as a function of order parameter along the physical trajectory: with crossover~(red), crossing CEP~(green) and with FOPT~(blue).}\label{fig:schematic}
\vspace*{-1mm}
\end{figure}

Considering the general phase transition theory, one may recall that the evolution of the Landau thermodynamic potential in terms of the order parameter is distinct along the process crossing the FOPT or crossover as depicted in the schematic diagram in Fig.~\ref{fig:schematic}.
For the trajectory with a crossover, the Landau potential has only one minimum as a function of the order parameter,
that gradually shifts as the temperature and chemical potential changes.
For a FOPT, the thermodynamical potential shows  a behavior with two local minima, and the state (phase) jumps from one minimum to another during the phase transition. Note that the potential is always continuous as a function of the order parameter. Such a picture may also explain the validity of the  multiplicity distribution for the FOPT~\cite{Adam:2019,Koch:2021}.  The potential is in some sense a holistic observable for the phase transition, and
if the system experiences the CEP or FOPT during the collision, the phase transition  will  leave   marks on the potential even down to the freeze-out point.
In this work we then propose novelly a method to construct the thermodynamical potential from the experimentally observed baryon fluctuations at different collision energies.

{\it Constructing the thermodynamical potential from the fluctuations---}\label{sec:theory}
The Landau thermodynamical potential is generally a Taylor expansion in terms of the order parameter.
There are various choices for the order parameter that can describe the chiral phase transition of QCD.
Here to associate with the experimental observables directly, we take the baryon number density as the order parameter since it is distinct in the chiral symmetric phase and the chiral symmetry broken phase and is capable of describing the order of phase transition~\cite{McLerran:1987pz,Xin:2014DSE}.
The Landau thermodynamical potential can be directly written as:
\begin{gather}
  \Phi[n_{B}] =  \sum_{k} \frac{1}{k!}\alpha_{k} \left(n_{B} - \bar{n}\right)^{k} \, ,
\end{gather}
where the coefficient $\alpha_{k} = \frac{\partial^{k} \Phi}{\partial n_{B}^{k}} \big |_{n_{B} =\bar{n}}$ is the $k$-th order Taylor coefficient at a ``virtual" number density $(n_{B} - \bar{n})$ with the physical number density $\bar{n}$
at given $(T,\bar{\mu})$.
The Landau potential can be related to the pressure by the Legendre transformation as $ \Phi[n_{B}] \equiv n_{B}[\mu_{B}^{}](\mu_{B}^{} -\bar{\mu}) - P[\mu_{B}^{}] $, where $\mu_{B}$ is the conjugate variable of the $n_{B}$ of the system, $P$ is the pressure.
At the physical point one has $\mu_{B}^{} = \bar{\mu}$, $n_{B}^{}[\bar{\mu}]=\bar{n}$ and $\Phi[\bar{n}] = - P[\bar{\mu}]$.
The potential's derivatives to  $n_{B}^{}$ can be expressed by the fluctuations order by order as:
\begin{gather}\label{eq:dPdnB}
 \alpha_{k} \! = \! \frac{\partial \alpha_{k-1}}{\partial n_{B}^{}}\Big{|}_{n_{B}^{}=\bar{n}_{0}} \!\! = \! \frac{1}{C_{2} }\frac{\partial \alpha_{k-1} \left[ C_{1}, \cdots , C_{k-1}; \mu_{B}^{} \right] }{\partial ({\mu_{B}^{}}/T)}\Big{|}_{\mu_{B}^{} = \bar{\mu}},   \\
 \alpha_{1}                         
 = (\mu_{B}^{} -\bar{\mu})\big{|}_{\mu_{B}^{} = \bar{\mu}}=0,
\end{gather}
with $C_{1} =\bar{n}$ and $C_{2} =\partial n_{B}/\partial (\mu_{B}/T) $, and in general the $k$-th order susceptibility $C_{k}$  reads:
\begin{equation}
  C_{k} = \frac{1}{T}\frac{\partial^{k} P}{\partial ({\mu_{B}^{}}/T)^{k} } \, .
\end{equation}
Since it is the order of the phase transition to be concerned here rather than the exact values,
we would like to redefine the potential with the dimensionless density order parameter $\tilde{n} = {n_{B}^{}}/\bar{n} - 1$ as:
\begin{equation}\label{eq:potential_Taylor}
\Omega[\tilde{n}]=\frac{\Phi[n_{B}]}{T\bar{n}}= \sum_{k=2}^{\infty} \omega_{k} \frac{\tilde{n}^{k}}{k!} \, ,
\end{equation}
with dimensionless Taylor coefficients:
\begin{eqnarray}\label{eq:Omegan}
& \omega_{2} = \frac{1}{ R_{21}}  ,\quad\quad\omega_{3} =   -\frac{R_{32}}{\, R_{21}^{2} } ,\qquad
 \omega_{4} = \frac{3\,R_{32}^{2} - R_{42}}{\, R_{21}^{3} }, \quad \;\; \notag\\
 &\omega_{5} = -\frac{15\,R_{32}^{3} - 10 R_{42}R_{32} + R_{52}}{\, R_{21}^{4} }, \qquad \qquad \qquad \qquad \qquad \\
 &\omega_{6} = \frac{105\,R_{32}^{4} - 105R_{42}R_{32}^{\,2} + 10R_{42}^{2} + 15R_{52} R_{32} -R_{62}}{\, R_{21}^{5}}, \;\; \cdots \, , \;\; \notag
\end{eqnarray}
where we have taken  $R_{ij} \equiv C_{i}/C_{j}$.
Therefore, the Landau potential can be determined completely by the observables of RHIC experiments.
Now if the Landau potential is capable of describing the FOPT, its Taylor expansion is required to be at least with 4-th order $\omega_{4}$,
and the monotonicity of the potential determines whether a FOPT happens.
The Landau potential is a monotonous function above (or below) the physical solution for the entire process of a crossover,
while for the FOPT, the coexistence of two phases exhibits that one local minimum degenerates with the physical one,
and a saddle point in the potential reveals the end-point of the coexistence region of the FOPT or the occurrence of the CEP.
Especially if one considers the fluctuations at the freeze-out point, it is in the hadron phase with a lower number density.
Therefore, one can set the criterion with having experienced a FOPT or passed through a CEP as that for $\tilde{n} \geq 0$, there exists a region with:
\begin{equation}\label{eq:criterion}
\frac{ \partial \Omega[\tilde{n}]}{\partial \tilde{n}} \leq 0 \, .
\end{equation}
Now  since the potential is a $k$-th order polynomial with a minimum at $\tilde{n}=0$, which implies:
\begin{equation}
\frac{ \partial \Omega[\tilde{n}]}{\partial \tilde{n}} = \tilde{n} \left( \frac{\omega_2}{1!} + \frac{\omega_3}{2!} \tilde{n} + \frac{\omega_4}{3!} \tilde{n}^2 + \cdots \right), \notag
\end{equation}
and the criterion of Eq.~(\ref{eq:criterion}) is therefore related to the discriminant $\Delta$ of the $(k-2)$-th order real coefficients polynomial $\varphi[\tilde{n}]$ after eliminating the root at $\tilde{n}=0$ from $\Omega[\tilde{n}]$ which can be denoted as:
\begin{equation} \label{eqm:Def-phi}
\varphi[ \tilde{n}]=\frac{1}{ \tilde{n}}\frac{ \partial \Omega[\tilde{n}]}{\partial \tilde{n}}= \sum_{k=0}^{\infty} \omega_{k+2}\frac{\tilde{n}^{k}}{(k+1)!}\,.
\end{equation}
The discriminant $\Delta$ of $ \varphi[ \tilde{n}]$  is  the determinant of the Sylvester matrix with $\varphi[\tilde{n}]$ and the first order derivative of $\varphi[\tilde{n}]$ up to a common factor~\cite{newstead1995discriminants}.  For the potential up to $O(\tilde{n}^4)$, the criterion is simple as it is the discriminant of the quadratic polynomial as:
\begin{equation}\label{eq:critwk}
  \Delta= \frac{1}{4} \omega_{3}^{2} - \frac{2}{3}\omega_{2} \omega_{4} \geq 0 ,
\end{equation}
in terms of the Taylor coefficients $\omega_{k}$.  One requires $\Delta \geq 0$  so that the potential contains two more extreme points besides the physical state at $\tilde{n}=0$,  one minimum for the meta-stable state and one maximum in between,
which is then the feature of a FOPT. The equality $\Delta= 0$ is reached at the end-point of the FOPT for the two roots becoming degenerate
or at the CEP for three roots becoming degenerate,
where one may further require the second derivative of the potential to be vanishing.
Note that the coefficient of the highest order should be positive for a stable system, as here, $\omega_{4} > 0$.
A negative $\omega_{4}$ means the higher order fluctuations are required.

\begin{figure*}[ht]
  \centering
  \includegraphics[width=0.20\textwidth]{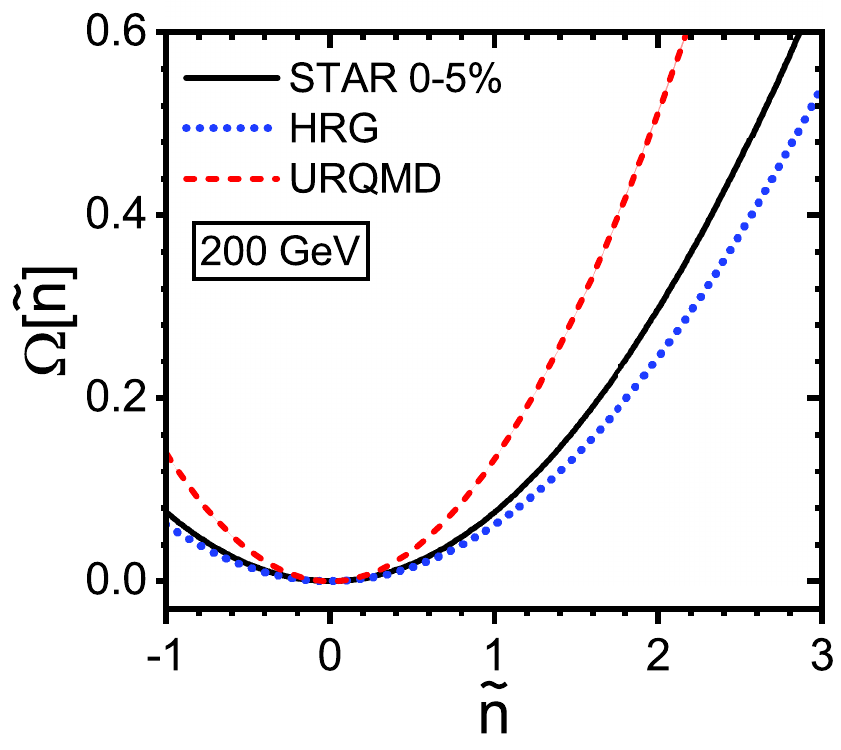}
  \includegraphics[width=0.18\textwidth]{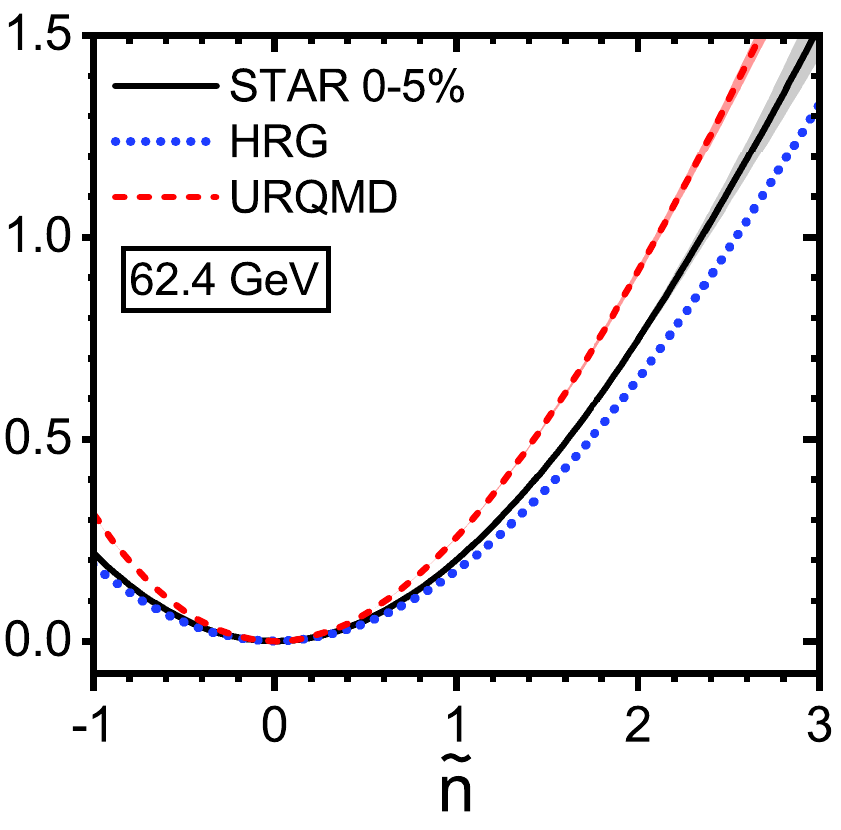}
  \includegraphics[width=0.18\textwidth]{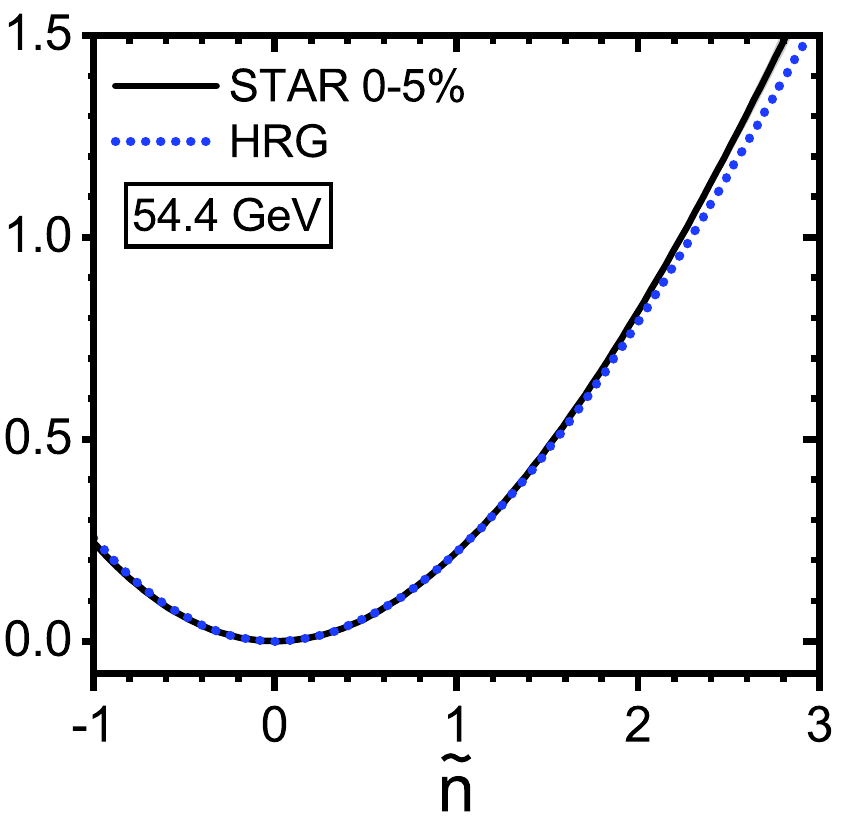}
  \includegraphics[width=0.182\textwidth]{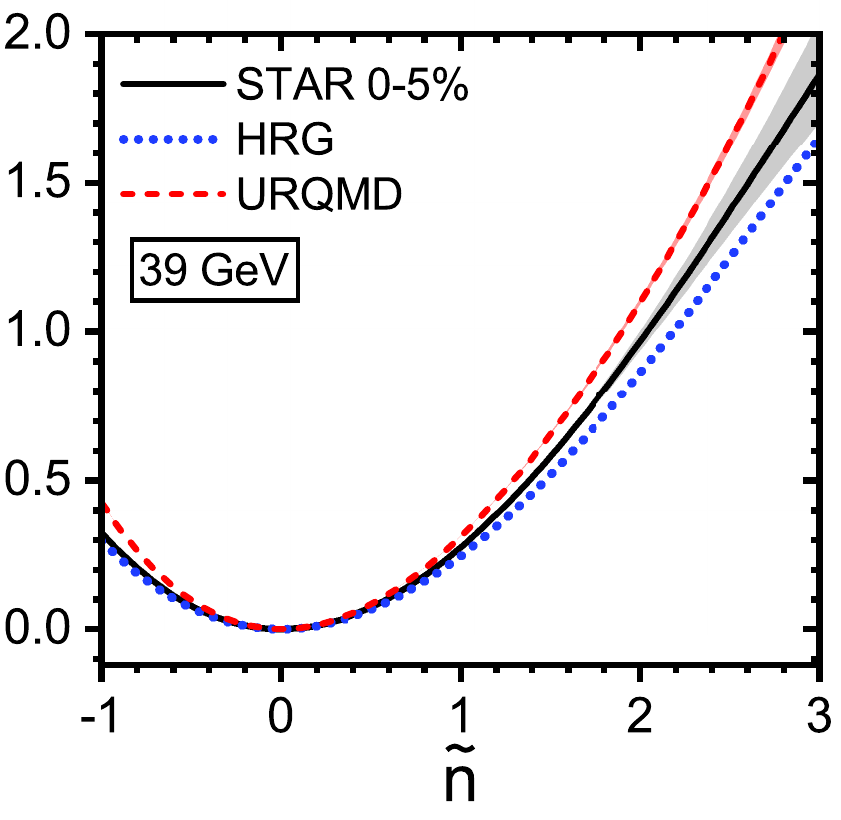}
  \includegraphics[width=0.172\textwidth]{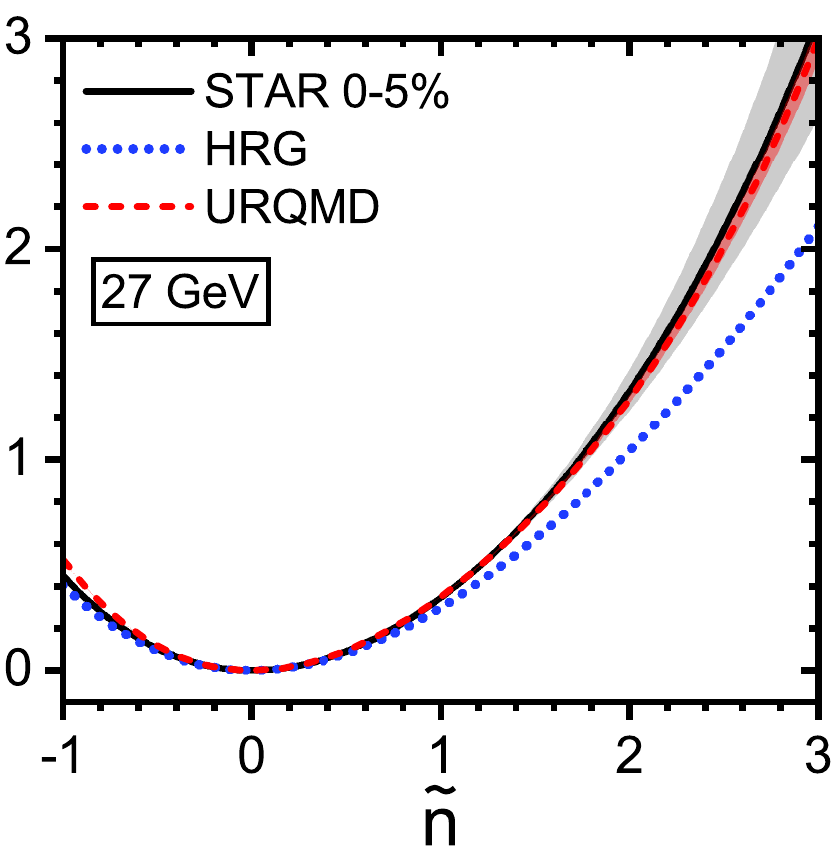}\\
  \includegraphics[width=0.20\textwidth]{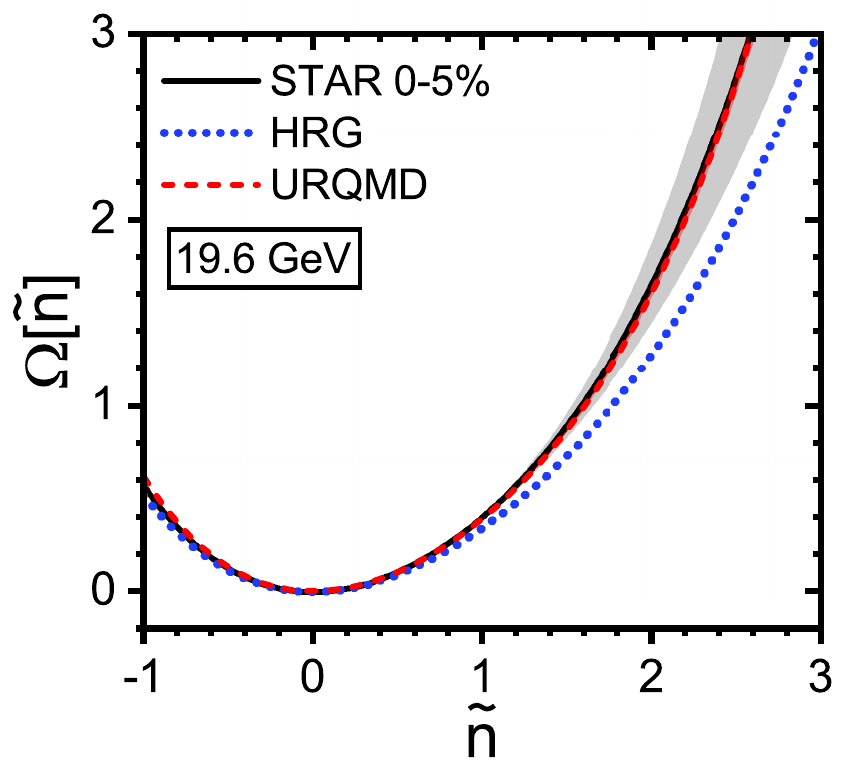}
  \includegraphics[width=0.18\textwidth]{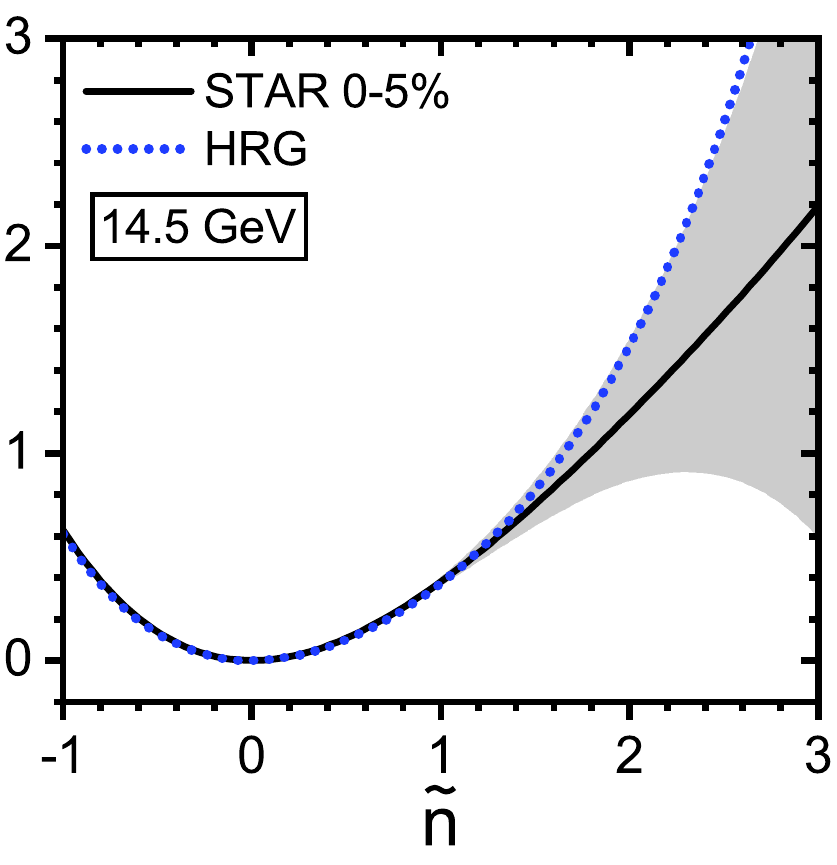}
  \includegraphics[width=0.18\textwidth]{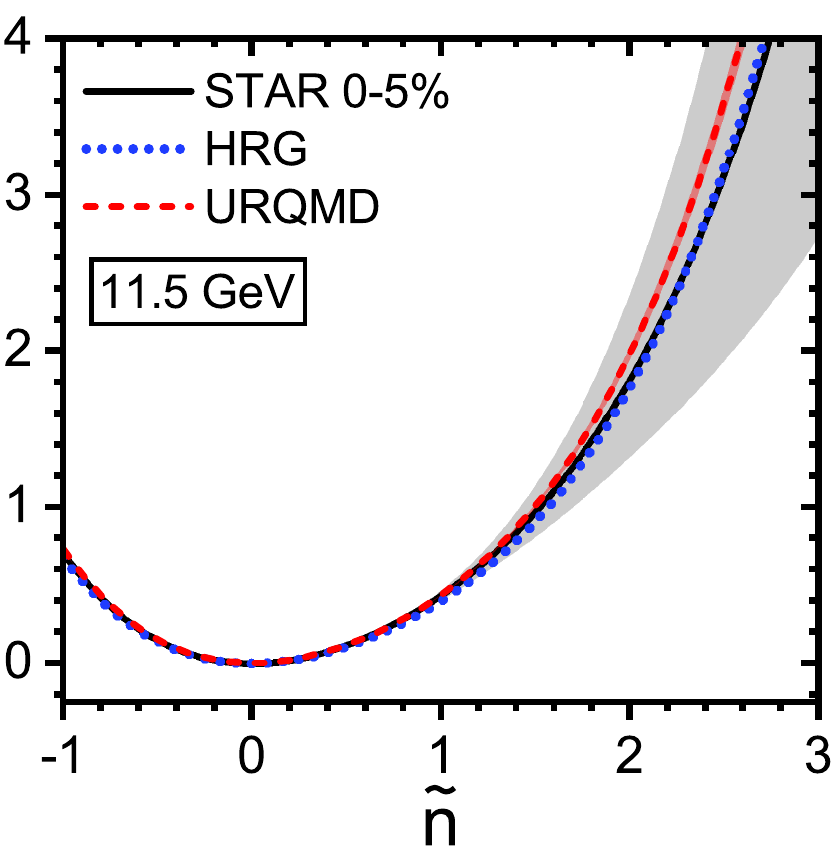}
  \includegraphics[width=0.18\textwidth]{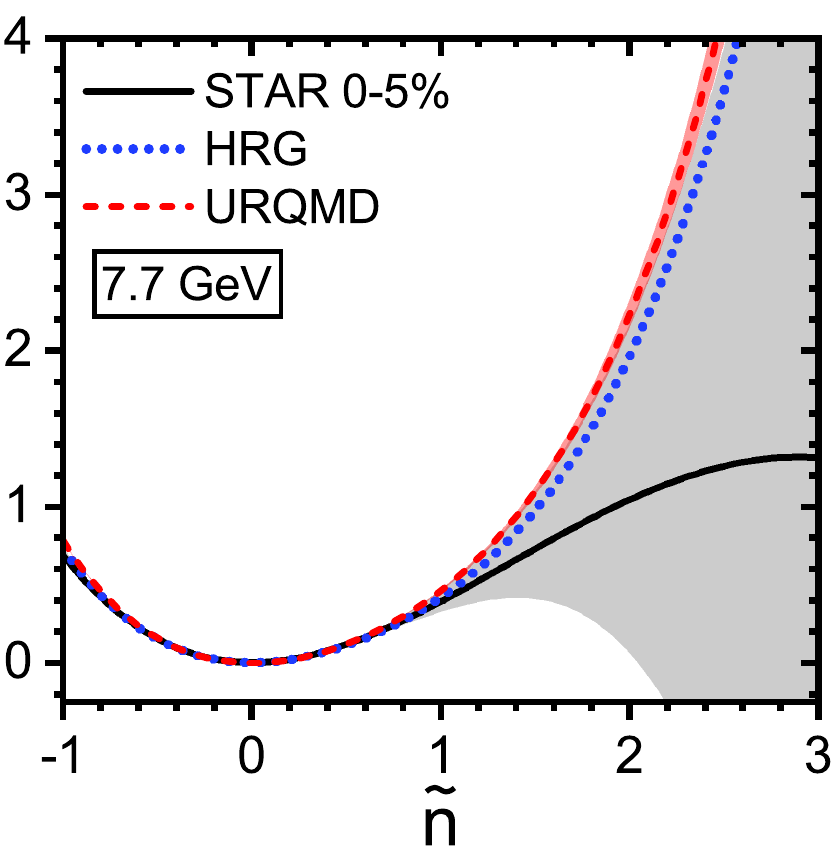}
  \includegraphics[width=0.18\textwidth]{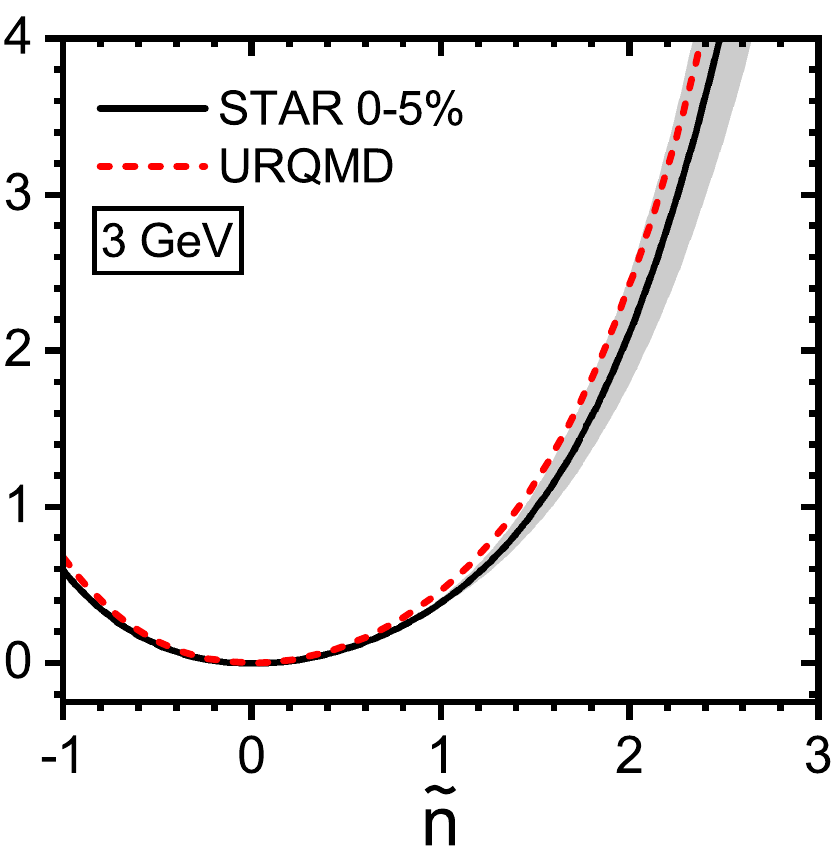}\\
\vspace*{-3mm}
\caption{(colored online) The obtained thermodynamical potential up to $O(\tilde{n}^{4})$ from the STAR data of the 0-5\% centrality at $\sqrt{s_{\textrm{NN}}} = 200$, 62.4, 54.4, 39, 27, 19.6, 14.5, 11.5, 7.7 and 3 GeV~\cite{STAR:2021kur,Abdallah:2022_3GeV} (the shadow area indicates the uncertainty range), and those from theoretical calculations with the HRG~\cite{Munzinger:2021HRG} and UrQMD~\cite{STAR:2021kur1}.
}   \label{fig:OmegaBES0_5}
\vspace*{-1mm}
\end{figure*}

One can also rewrite this criterion in the form of cumulant ratios:
\begin{equation}\label{eq:crit}
   \Delta=\frac{ 8 R_{42}-21 R_{32}^2 }{ 12 R_{21}^4} \equiv   \frac{8 \kappa \sigma^2 -21 (S\sigma)^2 }{ 12 (\sigma^2/M)^4}\geq 0,
\end{equation}
where $M$, $\sigma^{2}$, $S$, and $\kappa$ are the mean value, variance, skewness and kurtosis, respectively.
The cumulant criterion in Eq.~(\ref{eq:crit}) shows that for a FOPT, the high-order cumulant $C_4$ should be sufficiently larger than the lower-order ones (here they are $C_3$ and $C_2$), which is in agreement with the theoretical prediction on the critical scaling~\cite{Stephanov:2009,Stephanov:2011} and also verified by the observed skewness and kurtosis in Ref.~\cite{STAR:2021kur}.
With the criterion,  one may  extract  the possible signal of the CEP and the FOPT  from the experimental observables.
It also needs to mention that the criterion from the discriminant is stronger than the criterion in Eq.~(\ref{eq:criterion}),
as Eq.~(\ref{eq:criterion}) only requires the existence of one maximum.
However, if there is no other minimum,
Eq.~(\ref{eq:criterion}) will make the potential approach to negative infinity which leads to an unstable system.

For the potential up to $O(\tilde{n}^6)$,  one can also take the respective discriminant as the criterion.
The detailed formula can be found in the supplemental material.
For much higher order  polynomials, the discriminant is only a necessary condition for FOPT  and no  simple relation for the criterion,
but one can always check numerically if the criterion in Eq.(\ref{eq:criterion}) is satisfied.
The general discussions of the discriminant  and its relation with FOPT are also put in supplemental material.

\medskip

{\it Landau Potential with Experimental Data ---}
As the experiments have now reached the measurements till 4-th order fluctuations in the centrality of 0-5\% in a wide range of  collision energy from $200$~GeV down to $3$~GeV~\cite{STAR:2021kur,STAR:2021kur1,Abdallah:2022_3GeV},
the Landau potential in Eq.~(\ref{eq:potential_Taylor}) is thus capable of being constructed directly from the data.
The obtained results are illustrated in Fig.~\ref{fig:OmegaBES0_5}.

\begin{figure}[t]
\centering
\includegraphics[width=0.42\textwidth]{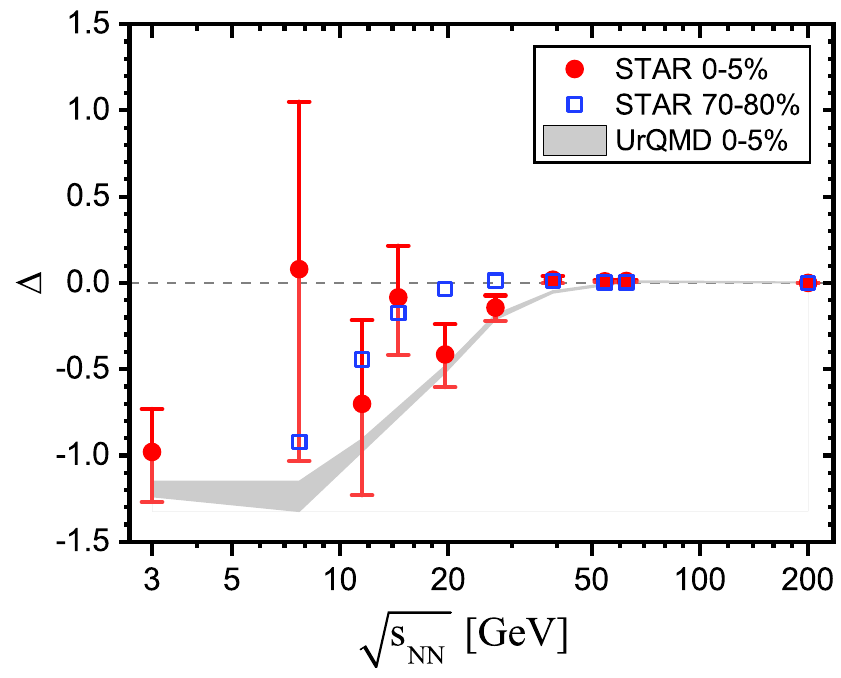}
\vspace*{-3mm}
\caption{(colored online) The  criterion  $\Delta$ for the FOPT from Eq.~(\ref{eq:crit}) with $0$-$5\%$ centrality collisions at $\sqrt{s_{\mathrm{NN}}^{}}=200$, $62.4$, $54.4$, $39$, $27$, $19.6$, $14.5$, $11.5$, $7.7$ and 3~GeV.
  The collision process has experienced a FOPT if $\Delta>0$.
  For high energy, the potential becomes quadratic and $\Delta$ is  infinitesimal because the number density is no longer efficient as the order parameter. The results for fluctuations with $70$-$80\%$  centrality are also depicted together with the results with UrQMD computation.
}\label{fig:criterion}
\vspace*{-1mm}
\end{figure}

\begin{figure*}[t]
    \centering
\includegraphics[width=0.304\textwidth]{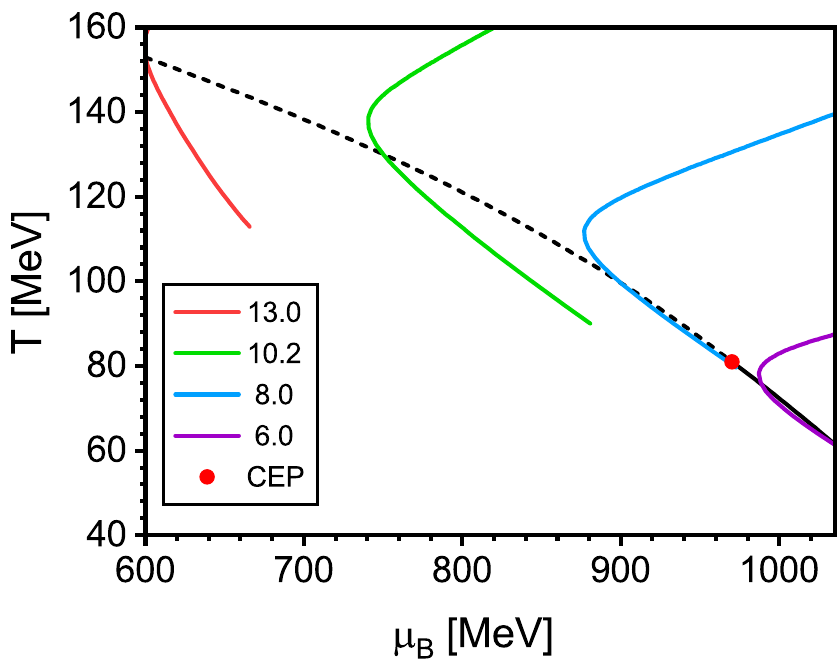} \hspace*{2mm} 
\includegraphics[height=0.24\textwidth]{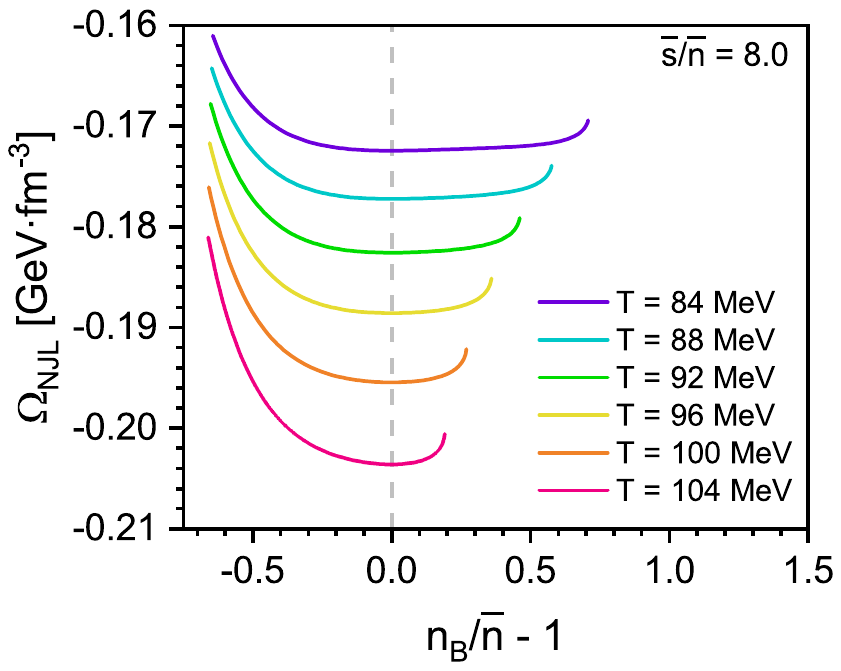}
\includegraphics[height=0.24\textwidth]{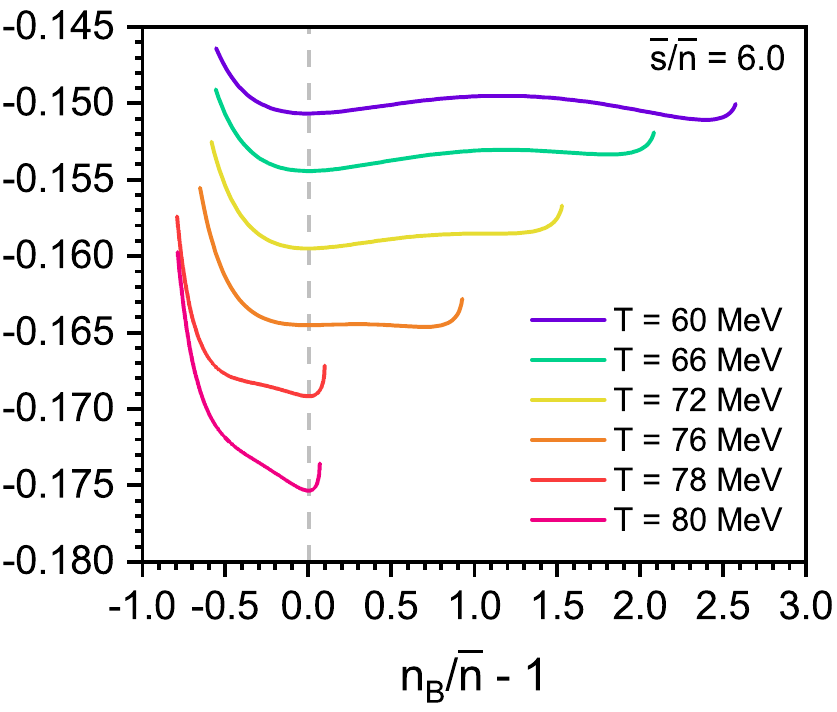}
\vspace*{-2mm}
\caption{(colored online)
Calculated phase diagram and several isentropic trajectories (labelled with the $\bar{s}/\bar{n}$ values) with crossover or FOPT via the NJL model ({\it left panel}),
and the parameter dependence of the potential $\Omega_{\textrm{NJL}}$ on the baryon density (order parameter) $n$ (rescaled by the physical one  $\bar{n}_{} = n[M_{\textrm{phy.}}]$ ) at given $(T,\mu_B)$ along two typical trajectories of crossover ($\bar{s}/\bar{n}=8.0$) and FOPT ($\bar{s}/\bar{n}=6.0$)  ({\it middle and right panels}). The $(T,\mu_{B}^{})$ position on each trajectory is controlled by the temperature.
%
%
}\label{fig:OmegaN}
\end{figure*}

As shown in Fig.~\ref{fig:OmegaBES0_5}, it is found that the data yield a one minimum (or monotonous in the region $\tilde{n} > 0$) potential for $\sqrt{s_{\mathrm{NN}}} \geq 11.5$ GeV and thus a smooth crossover phase transition.
Now it is interesting  that at  $\sqrt{s_\mathrm{NN}}=7.7$ GeV the potential changes its monotonicity in the region of $\tilde{n} > 0$,
which indicates that the corresponding trajectory may has crossed the phase boundary with a FOPT.  However,
due to the large statistical error in the experimental data, this non-monotonicity is still not robust and requires further experiments to confirm.

It is however slightly surprising that at  $\sqrt{s_{\mathrm{NN}}^{}}= 3$~GeV, the signature of FOPT disappears.
One possible reason is that the freeze-out point in the case of 3~GeV  is too far away from the phase transition line(s) so that the possible FOPT signature is wiped out, or even the collision starts below the phase transition line.

One can also demonstrate the criterion in Eq.~(\ref{eq:crit})  as in Fig.~\ref{fig:criterion}.
For $\sqrt{s_\mathrm{NN}^{}} > 39$~GeV, the discriminant is close to zero because the potential is mostly quadratic with $R_{21}$  being dominant.
There in general does not exist another extreme point of the potential for these collision energies as shown in Fig.~\ref{fig:OmegaBES0_5}.
Note that at  $\sqrt{s_\mathrm{NN}^{}}=39$~GeV, the value of the discriminant is slightly larger than zero, which is   mainly due to the error of the measurement for the third and fourth order fluctuations and also the missing of higher order fluctuations. Moreover, the quadratic behavior of the potential  also indicates  that at large collision energy, the number density is no longer efficient as the order parameter since it becomes indistinguishable for the two phases.

For lower collision energy, the potential becomes quartic and enables the possibility of a FOPT.
At $\sqrt{s_{\mathrm{NN}^{}}} = 14.5$~GeV,  only the upper bound exceeds zero which respects the lower bound in the potential in Fig.~\ref{fig:OmegaBES0_5}.
Now one may look into the potential at $\sqrt{s_\mathrm{NN}^{}}=7.7$ GeV.
Though there is still large uncertainty as the error of the data is large,
the mean value of the experimental data clearly satisfies the criterion.
This is consistent with the previous finding from Fig.~\ref{fig:OmegaBES0_5}. It also shows the sensitivity of the criterion, which makes it  possible to reveal the signal of FOPT   when equipping with   the higher precision data.    Note that   the error of $C_{4}$  is dominant in the error band of  the potential, and hence, it is essential to improve the measurement of $C_{4}$ in comparison to the other fluctuations.

Besides, the cumulant ratios from some theoretical calculations are also incorporated in Fig.~\ref{fig:OmegaBES0_5} and ~\ref{fig:criterion},
including those with the hadron resonance gas~(HRG) model~\cite{Munzinger:2021HRG} and the UrQMD transport model~\cite{Bleicher:1999}.
We verify that the potential constructed from the fluctuations of the HRG and UrQMD always give a monotonous behavior in the $\tilde{n} > 0$ region with also the discriminant always being negative,
which is consistent with the fact that there is no FOPT in these approaches.
The results for larger centrality  in the $70-80\%$ regions are also shown in Fig.~\ref{fig:criterion}.
The criterion for FOPT is not satisfied for all energies in case of such a centrality collisions.
Incorporating with the theoretical methods like DSE and fRG which have predicted the existence of the CEP may be able to give a proper estimation at which collision energy the FOPT can be observed from the fluctuations~\cite{Isserstedt:2019pgx, Fu:2021oaw, Gao:2021vsf},
and the work is under progress.
%

\medskip

{\it  An illustration with  effective model calculation---}\label{sec:validation}
To verify the above results obtained from general analysis and experimental data,
we illustrate the complete evolution procedure of the thermodynamic potential during the physical process,
which is obtained with the simple Nambu--Jona-Lasinio~(NJL) model~\cite{Buballa:2005njl}.
For a degenerate $N_{f}$ flavor system, the potential in the NJL model reads:
\begin{gather}
\nonumber  \frac{\Omega_{\textrm{NJL}}[M]}{V} = \frac{(M-m_0)^2}{4G} - 2 N_c N_f \int_0^\Lambda \frac{d^3 \boldsymbol{p}}{(2 \pi)^3} \bigg\{ E_{\boldsymbol{p}}  \qquad \quad  \\
\;  + T\,\log\left[1\! +e^{-(E_{\boldsymbol{p}}-\mu)/T} \right] + T\,\log\left[ 1 \! +e^{-(E_{\boldsymbol{p}}+\mu)/T} \right] \bigg\},   \label{eq:OmegaNJL}
\end{gather}
where $M$ is the mean-field constituent quark mass which is also a typical  order parameter of the chiral phase transition, $m_0$ is the current quark mass, $\mu = {\mu_{B}^{}}/3$ is the degenerate quark chemical potential, $E_{\boldsymbol{p}} = \sqrt{\boldsymbol{p}^{2} + M^{2}}$ and $\Lambda$ the momentum cut-off of the model.

The physical state corresponding to the potential are determined by the gap equation $\delta \Omega/ \delta M = 0$, which reads explicitly as
\begin{align}\label{eq:gapeqNJL}
\nonumber M & ^{\textrm{phy.}} = m_0 + N_c N_f \, G M^{\textrm{phy.}} \int_0^\Lambda \frac{p^2\,dp}{\pi^2\,E_p^{\textrm{phy.}}}  \\
& \times  \left[ \tanh\left(\frac{E_p^{\textrm{phy.}}-\mu}{2T}\right) + \tanh\left(\frac{E_p^{\textrm{phy.}}+\mu}{2T}\right) \right].
\end{align}
For a stable state, the additional condition $\delta^{2} \Omega/ \delta M ^{2} > 0$ is also implemented so that it locates at a local minimum of the potential.

The phase diagram can be then calculated. For the FOPT,
the phase transition boundary is determined with the criterion that the potential of the Nambu phase is equal to that of the Wigner phase~\cite{Jiang:2013njl,Xin:2014pNJL}, $\Omega[M_{N}] = \Omega[M_{W}]$.
For the crossover, the phase boundary is determined by the maximum of the chiral susceptibility $\chi_{T}^{} = \partial M/\partial T$.
We consider the case with $N_{f} = 2$ light flavors and implement the model parameters: $m_{0} = 5.6$~MeV, $\Lambda = 587.9$~MeV
and $G\Lambda^{2} = 2.44$.

We then consider the isentropic trajectories  with $\bar{s}/\bar{n}=constant$ as the physical evolution trajectories. $\bar{s}$ and $\bar{n}$ are the entropy and the baryon density at the physical point with $M=M^{\textrm{phy.}} $, and as usual, the entropy $s$ and number density $n_{B}^{}$ is determined from the potential in Eq.~(\ref{eq:OmegaNJL}) as
\begin{eqnarray}\label{eq:nBnphyNJL}
s[M; T, \mu_{B}^{}]& \displaystyle = -\left(\frac{\partial\, \Omega[M;T,\mu_{B}]}{\partial\, T}\right)_{M,T},\\
n_{B}[M; T, \mu_{B}^{}]& \displaystyle = -\left(\frac{\partial\, \Omega[M;T,\mu_{B}]}{\partial\, \mu_{B}}\right)_{M,T}. \label{eq:nBnphyNJL1}
\end{eqnarray}
The obtained results of the phase diagram and several isentropic trajectories are shown in Fig.~\ref{fig:OmegaN}.

It is now interesting to look through the evolution of the thermodynamical potential along the trajectories.
If regarding the mass $M$ as  the order parameter, which has been studied widely (e.g. Refs.~\cite{Buballa:2005njl,Xin:2014pNJL}), one found that the evolution behavior is closely related to the QCD phase structure.
Since we are interested in the virtual baryon density dependence which is related to the cumulants ratios,
we convert the original order parameter $M$ to baryon density $n[M]$ by the relation in Eq.~(\ref{eq:nBnphyNJL1}),
which satisfies $\bar{n} = n[M_{\textrm{phy.}}]$ at the physical point.

The parameter dependence of $\Omega[n_{B}]$ is computed along each certain trajectory
in a temperature range $T - T_{c} \in (-15,+5)$~MeV with $T_{c}$ the critical one at which the trajectory intersects with the phase boundary.
Here we pick two benchmarks $\bar{s}/\bar{n}= 6.0$, 8.0 for the FOPT, the crossover, respectively.
The results are also shown in Fig.~\ref{fig:OmegaN}.
It is found evidently that, within such a range of $\;T - T_{c}$, the $\Omega[n_{B}]$ in case of the FOPT is in general non-convex and non-monotonous near the phase boundary,
whereas in the crossover region, e.g., the case with $\bar{s}/\bar{n} = 8.0$, the $\Omega[n_B]$ is always convex even if the trajectory may travel very close to the CEP.
In short, the thermodynamical potential is distinct from each other in the cases of the crossover and the FOPT.
Such a feature maintains at least within the temperature range $T - T_{c} \in (-15,+5)$~MeV along the trajectories, so that it is possible to survive in the experimental measurement~\cite{Fu:2020,Gao:2020prd}.

{\it Summary---}
The thermodynamical potential opens an access to track the physical process in the heavy-ion collision,
and thus makes it possible to probe the signal of the CEP or FOPT even it occurs at the freeze-out point,
which is away from the phase boundary.
The criterion from the potential is more general and powerful than that from the fluctuation itself because of the synergy that the potential assembles all orders of the fluctuations to the verdict of the signal of phase transition.
Especially, if the potential can be constructed in a model-independent way, it would be greatly helpful  to search for the phase transition experimentally.
Therefore, We  proposed a novel method to construct the potential directly through the experimental data in this Letter. With the net-proton fluctuations measured in Au+Au collisions at $\sqrt{s_\mathrm{NN}}$ = 3--200 GeV by the STAR experiment, we obtain the Landau potential at each energy and the energy dependence of the criterion for the FOPT. It is found that the criterion derived from the fluctuations from the peripheral (70-80\%) Au+Au collisions and UrQMD model are negative and show monotonic decreasing as a function of energy, indicating no signal of FOPT. Further, the criterion are negative for almost all the energies with the 0-5\% most central data and   consistent with the results from UrQMD with large uncertainties. An exception is that, at $\sqrt{s_\mathrm{NN}}=7.7$ GeV, the mean value of the current data satisfies the criterion of FOPT.
However, the uncertainties of the experimental measurement are still too large to give a conclusive assertion.
Besides, though the quadratic potential is already enough for describing the FOPT,
the higher order fluctuations can help to confirm the convergence of the Taylor expansion.
In all, a high precision measurement of the $C_{1,...,6}$ in 0-5\% centrality in the neighborhood of  $\sqrt{s_\mathrm{NN}}=7.7$ GeV is of primary importance to finally verify the FOPT of QCD.

\section{Ackowledgement}
The authors thank the fQCD collaboration~\cite{fQCD:2022}  for fruitful discussions. YL and YXL was supported by the National Natural Science Foundation of China (NSFC) under Grant Nos. 11175004 and 12175007.
XL was supported by the National Key Research and Development Program of China with grant Nos. 2022YFA1605501, 2020YFE0202002 and 2018YFE0205201,
the NSFC with grant Nos. 12122505 and 11890711) and the Fundamental Research Funds of the Central China Normal University with grant No. CCNU220N003. LC was supported by the NSFC with Grant No. 12135007.

\bibliography{Ref-potential}

\vspace*{4mm}



 \section{Supplemental material: The criterion of CEP and FOPT and the discriminant of the polynomial}

Considering the criterion function $\varphi[x]$  as a general $n$-th order polynomial of argument $x$ as:
\begin{equation}
\varphi[x] = a_{n} x^{n} + \cdots + a_{0} ,
\end{equation}
with $\displaystyle a_{n} = \frac{\omega_{n+2}}{(n+1)!}$,
which is a polynomial with two degrees lower than the potential $\Omega$ (as defined in Eq.~(\ref{eqm:Def-phi})).
If the potential describes the FOPT, it must have two minima and one maximum.
One minimum is the physical state and the other minimum is the meta-stable state.
Therefore,  $\varphi[x]$ should contain two real roots since the root for physical state has been eliminated.
The property of the roots is in general related to its discriminant $\Delta$ which is defined as:
\begin{equation}
\Delta=(-1)^{\frac{1}{2}n(n-1)}\frac{1}{a_{n}}R
\end{equation}
with $R$ the determinant of its respective $(2n-1)\times(2n-1)$  Sylvester matrix as:
\begin{widetext}
\begin{equation}
\label{eq:matrix}
\begin{gathered}
\begin{bmatrix}
a_{n} &  a_{n-1} & a_{n-2} &\cdots & a_{1} & a_{0} & 0 & \cdots & 0  \\[1mm]
0 &  a_{n} &a_{n-1} & a_{n-2} & \cdots & a_{1} & a_{0} & \cdots& 0  \\[1mm]
\vdots & \vdots & \vdots & \vdots & \vdots & \vdots & \vdots & \vdots &\vdots  \\[1mm]
 0 & 0 & 0 & a_{n} &a_{n-1} &  a_{n-2}  & \cdots & \cdots & a_{0} & \\[1mm]
n a_{n} &  (n-1) a_{n-1}  & (n-2) a_{n-2} & \cdots& a_{1} & 0 & \cdots & \cdots & 0  \\[1mm]
0 & n a_{n} &  (n-1) a_{n-1} & (n-2) a_{n-2} & \cdots& a_{1} & \cdots & 0 & 0  \\[1mm]
\vdots & \vdots & \vdots & \vdots & \vdots & \vdots & \vdots & \vdots & \vdots  \\[1mm]
 0& 0& \cdots & 0 & n a_{n} &  (n-1) a_{n-1}  & \cdots & \cdots & a_{1} & \\
\end{bmatrix}
\end{gathered}
\end{equation}
\end{widetext}

Generally speaking, for $n$-th order polynomials, one requires two real roots in $\varphi[x]$ to describe the feature of a FOPT, and the surplus roots should be complex in order to avoid more minima in the potential which are not physical.  Now for polynomials with real coefficients as considered here, the complex roots can only exist in pairs and moreover,  there exist even-number pairs of complex roots with  the discriminant $\Delta>0$,  while  exist odd-number pairs for $\Delta<0$.
Therefore, for an odd-order potential, the related $\varphi[x]$ then has an odd number of real roots, rather than two. Also, despite the fact that Eq.~(\ref{eq:criterion}) can be satisfied in this case, the potential is unstable for an infinitely large density as it diverges to negative infinity.
The order of $\varphi[x]$ should be even to guarantee that it has exactly two real roots.
For the case of order $2(2m+1)$, one entails $\Delta>0$ which brings in even-number pairs of complex roots and thus satisfies the requirement of containing two real roots,  while for order $2(2m)$, one then entails  $\Delta<0$.
Note that the discriminant does not constrain completely the number of the roots, the detailed technique is beyond the scope of this work and here we simply apply the discriminant as the criterion which completely constrains the polynomials to have two real roots  in the quadratic and quartic cases.

For a fourth order potential, $\varphi[x]$ is a quadratic polynomial and thus one has the criterion for having CEP and FOPT as:
\begin{equation} \label{eq:cri4}
  \Delta = a_{1}^{2} - 4a_{2} a_{0} = \frac{1}{4} \omega_{3}^{2} - \frac{2}{3}\omega_{2} \omega_{4} \geq 0 \, ,
\end{equation}
which is just the criterion in Eq.~(\ref{eq:critwk}).

For sixth order potential, the discriminant  is lengthy. We simply put the criterion in $a_{k}$ here as:
\begin{eqnarray}\label{eq:cri6}
  \Delta & = &  a_{1}^{2} a_{2}^{2} a_{3}^{2} - 4 a_{0} a_{2}^{3} a_{3}^{2} - 4 a_{1}^{3} a_{3}^{3} + 18 a_{0} a_{1} a_{2} a_{3}^{3} \notag \\[1mm]
 & & - 27 a_{0}^{2} a_{3}^{4} - 4 a_{1}^{2} a_{2}^{3} a_{4} + 16 a_{0} a_{2}^{4} a_{4} + 18 a_{1}^{3} a_{2} a_{3} a_{4} \notag \\[1mm]
 & & - 80 a_{0} a_{1} a_{2}^{2} a_{3} a_{4} - 6 a_{0} a_{1}^{2} a_{3}^{2} a_{4} + 144 a_{0}^{2} a_{2} a_{3}^{2} a_{4}   \notag \\[1mm]
 & & - 27 a_{1}^{4} a_{4}^{2} + 144 a_{0} a_{1}^{2} a_{2} a_{4}^{2} - 128 a_{0}^{2} a_{2}^{2} a_{4}^{2} \notag \\[1mm]
 & & - 192 a_{0}^{2} a_{1} a_{3} a_{4}^{2} + 256 a_{0}^{3} a_{4}^{3} \leq 0\, .
\end{eqnarray}
When $\omega_{5}$ and $\omega_{6}$ are small, the above criterion in Eq.~(\ref{eq:cri6}) degenerates into Eq.~(\ref{eq:cri4}).

\end{document}